\documentclass [12 pt] {article}
\usepackage {amsmath}
\usepackage{amsthm, amssymb}
\usepackage {graphicx}
\usepackage{url}
\usepackage{fullpage}
\usepackage{multicol}
\usepackage{hyperref}
\usepackage{subfig}
\usepackage{xcolor}

\newtheorem{lem:upper_bound_g}{Lemma}[section]
\newtheorem{lem:lower_bound_L}[lem:upper_bound_g]{Lemma}

\newtheorem{thm}{Theorem}[section]

	\def\Bd{{\bf d}}

	\def\Bn{{\bf n}}

	\def\Bx{{\bf x}}

	\def\B0{{\bf 0}}

	\def\calD{{\mathcal D}}

	\newcommand{\GD}{\Delta}

	\def \calD{{\mathcal D}}

	\def \RR {{\mathbb R}}
	
	\def \ba {\begin{array}}
	\def \ea {\end{array}}

	\def\Bd{{\bf d}}

	\def\Bn{{\bf n}}

	\def\B0{{\bf 0}}
	
	\def\Bx{{\bf x}}

\title {\bf \Large Personal Sound Zones and Shielded Localized Communication through Active Acoustic Control}
\author{Neil Jerome A. Egarguin$^{1}$, Daniel Onofrei$^2$ \\ \small{$^1$Institute of Mathematical Sciences, University of the Philippines Los Ba\~nos}\\  \small{College, Laguna, Philippines}\\  \small{$^2$Department of Mathematics, University of Houston, Houston, TX, USA} \\ }

\date{}

\begin {document}

\maketitle

\begin{abstract}
In this paper, we present a time domain extension of our strategy on manipulating radiated scalar Helmholtz fields  and discuss two important applied scenarios, namely (1) creating personal sound zones inside a bounded domain and (2) shielded localized communication.  Our strategy is based on the authors' previous works establishing the possibility and stability of controlling acoustic fields using an array of almost non-radiating coupling sources and presents a detailed Fourier synthesis approach towards a time-domain effect. We require that the array of acoustic sources creates the desired fields on the control regions while maintaining a zero field beyond a larger circumscribed sphere. This paper recalls the main theoretical results then presents the underlying Fourier synthesis paradigm and show, through relevant simulations, the performance of our strategy. 

\end{abstract}

\section{Introduction}
\label{intro}

The research presented in this paper aims at investigating novel modalities for the characterization of feasible acoustic sources (i.e., position and feed inputs for active radiators) towards the active manipulation of their radiated fields in prescribed exterior regions and/or desired far field directions, with applications in localized communication and the realization of personal sound zones. There exists a vast body of literature on computational and analytical methods for acoustic scattering and radiation, such as the major monographs \cite{Scatering1, Martinbook}. 

Regarding the problem of active manipulation of acoustic fields, two of the earliest works concern the active control of sound, i.e., \cite{Leug} (feed-forward control of sound) and \cite{Olson} (feedback control of sound). The general problem of active control of scalar Helmholtz fields is discussed in  \cite{Popa, Guicking, Loncaric2001, Loncaric2005, Loncaric2004, Loncaric2003, Peterson2007, Fuller, peake, Elliot1990, Elliot2001, Ahrens2010, Acousticcontrol1, Miller2006, Vasquez2011, Vasquez2009, Loncaric2001, Loncaric2005, Han2018, Han2019, Menzies2012, Zhang2016, Zhang2019}. Energy localization results for Helmholtz scalar fields obtained via Runge approximations are discussed for the interior problem in \cite{EMHelmholtzfocusing} and for the scattering problem in \cite{EMfocusing4, EMfocusing5, EMfocusing6}, ( see also recent thesis \cite{Annalena23} and the study in \cite{Arens23} for localized wave functions in a waveguide).

The applications of acoustic field control include active noise cancellation (\cite{Nelson1992, Bonito1999, Cheer2015, Kim2017, Mao2018}), sound synthesis and reproduction (\cite{Ahrens2010, Jin2013, Omoto2015, Zhang2019, Galvez2019, Lee21, Gallian21, Gao23}), and active control of acoustic scattered fields with application to cloaking and shielding (\cite{Acousticcontrol2, Acousticcontrol6, Bobrovnitskii2010, Vasquez2011, Vasquez2009, broadband_vasquez, Liu2019, Egglera2019, Majid2018}). A more comprehensive discussion and analysis of some of the methods employed in these applications is also given in \cite{Cheer2016} for scattered field control, in \cite{Kuperman2011, Keller1997, MarineAcoustics, boyles} for direct problems in ocean and layered media and in \cite{Elliot2012, WFS1, Acousticcontrol3, Acousticcontrol4, Acousticcontrol5, Acousticcontrol7}, for general sound field control.

Boundary integral operators were used in \cite{Onofrei-S, Onofrei2014} to produce a stable unified control strategy in the case of a single active surface source proving the active control of scalar radiated fields in prescribed exterior region of space. These theoretical results were later extended by the authors' group to consider active scalar sources in ocean environments, free space or layered media, and applied  to relevant 2D and 3D numerical studies in \cite{Hubenthal2016, Platt2018, Egarguin2018, EgarguinWM2020, EgarguinIP2020, EgarguinAA2021, EgarguinIPSE2021}. In \cite{OnofreiAMS22}, the proposed active manipulation methods were extended and applied to the case where impenetrable obstacles were assumed present as well.

Practical implementation of such strategies may require accurate field measurements, or the use of non-traditional sources. In this regard and in the context of scattered field active control, the authors in \cite{Loncaric2001, Cheer2019, Acousticcontrol1} proposed methods which are only based on measurements of the total field around the sources, while in a more general context, in \cite{Andrew2019, Activecontrolthesis2022} the authors proposed methods to overcome the difficult task of using monopolar sources in the control schemes proposed in the literature. Nanometer size sources are recently being proposed in the literature based on strong variations of thermal effects in special materials nano-tubes (\cite{Brown2016, Qiao2020, Hu2014}).

The problem of field synthesis requires the construction of feed inputs on the active sources for the approximation of a given far field pattern \cite{Kirsch,Ahrens2012} (``the far field synthesis") (see also the monograph \cite{Devaney3} where general radiation theory and source synthesis techniques are discussed). 
The problem of field focusing is mainly formulated in terms of power maximization in the far field (with eventual nulls in prescribed far field directions) and analyzed through an associated eigenvalue problem stemming from first order optimality conditions imposed to an associated augmented Lagrangian (see the monograph \cite{Kirsch}). 

The problem of using active sources for creating highly localized patterns in prescribed regions of space with specified fields in given far field directions in general media is of potential high interest for many possible applications (see \cite{EMfocusing3}, \cite{EMfocusing4}, \cite{EMfocusing5} and \cite{EMfocusing6} for scattering energy localization results with applications to inverse obstacle problems as well as \cite{Annalena23} and \cite{Arens23}). 

Regarding the problem of time domain active field manipulation, we studied in \cite{Egarguin2018} the scalar case and by using superposition we produced simulations suggesting the possibility of using our scalar control theory for the characterization of causal sources so that their radiated fields approximate a given time domain signal in a near field region while maintaining a very small field in another exterior region.

In what follows, we will use our theoretical results on the active manipulation of fixed frequency Helmholtz fields reported in \cite{Onofrei2014, Hubenthal2016, Platt2018, Egarguin2018, EgarguinWM2020, EgarguinIP2020, EgarguinAA2021, EgarguinIPSE2021} to build a detailed Fourier synthesis approach towards time domain active control as applied to two application scenarios: 

\begin{itemize}
	\item the realization of personal sound zones inside a bounded domain; and \vspace{-0.8cm}\\
	\item shielded localized communications, i.e., design of an array of sources with a personal listening area and a fast decaying acoustic field.
\end{itemize}
Section \ref{prob} presents the problem in mathematical parlance, while Section \ref{math} and Section \ref{numerical} respectively discuss the mathematical and numerical frameworks of the proposed solution scheme. Section \ref{simulations} shows the implementation of the solution scheme through several numerical simulations of the scenarios mentioned above. The section includes a detailed formulation of the time-domain synthesis via Fourier analysis. The paper ends with some concluding remarks in Section \ref{conclusion}.

\section{Statement of the Problem}
\label{prob}

 Mathematically, the general question is to find acoustic pressure $p_j$ or normal velocity $v_j$, $j=\overline{1, n}$ on the surface of $n$ acoustic sources $D_j$, $j= \overline{1,n}$, so that the solution $u$ of the Helmholtz problem in the exterior of the sources given by
\begin{equation}
\vspace{0.15cm}\left\{\vspace{0.15cm}\begin{array}{llll}
\GD u+k^2u=0 \mbox{ in } R\!\setminus \left ( \displaystyle \bigcup_{j=1}^n \!\overline{D}_j \right ) \vspace{0.15cm},\\
\nabla u\cdot \Bn_{j}=v_j, (\mbox{ or } u=p_j)\mbox{ on }\partial D_j\vspace{0.15cm}, ~j= \overline{1,n}\\
\text{suitable condition on } \partial R
\end{array}\right.
\label{helmholtz1}
\end{equation}
approximates some predetermined fields $f_p$ on $m$ control regions $\displaystyle R_p\Subset R \setminus \left ( \bigcup_{j=1}^n \!\overline{D}_j \right )$, $p = \overline{1, m}$ interior to $R$, where the region $R$ could be a bounded domain encompassing the sources and the control regions or the whole space ${\mathbb  R}^3$ in which case the boundary condition in \eqref{helmholtz1} above becomes the classical Sommerfeld outgoing condition at infinity. 

\begin{figure}[!h]
\centering
\includegraphics[width=0.4\textwidth]{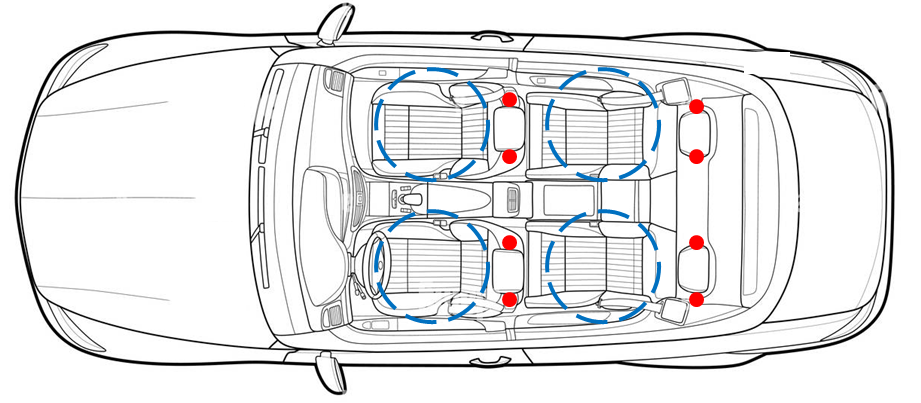}
\caption{Personal Sound Zones in a Car Cabin: In this application, we create sound profiles in the control regions inside the blue spheres using the speakers drawn as red dots}
\label{CarCabin}
\end{figure}

In our first application we consider the creation of personal sound zones in a car cabin, i.e., where $R$ is the region of the car interior. Figure \ref{CarCabin} shows a typical car cabin where we wish to generate predetermined sound profiles in the vicinity of each passenger, i.e., the control regions inside the blue spheres using speakers (sound sources) represented by red dots located in the headrest of each seat. This will allow each passenger to enjoy their acoustic preference with minimal to no interference with each other. One of the main difficulty with this application, besides the active control of the sources to create personalized sound experiences on each of the control regions, is the effect of the boundary of $R$. In the case of a car cabin, the boundary is composed of different materials such as glass, hard plastics, metal, leather, etc. The different reaction of these materials to sound makes the boundary condition in \eqref{helmholtz1} difficult to model. Moreover, the boundary condition on $\partial R$ may play a crucial role in the propagation of acoustic waves as it determines the reverberations of sound created by the sources back into the cabin. 

\begin{figure}[!h]
	\centering
	\includegraphics[width= 0.7 \textwidth]{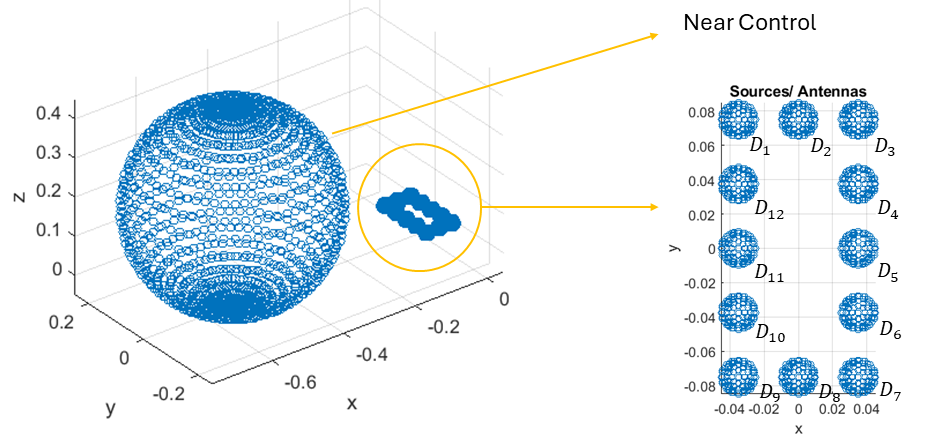}
	\caption{Shielded Localized Communication: In this application, we create a desired sound profile in the near control using the sources while requiring the acoustic field to decay fast}
	\label{PhoneNearGeom}
\end{figure}

The second application we consider is that of shielded localized communication, i.e., where $R$ denotes the whole space. In this case, the goal is to realize a free space communication zone shielded from the exterior by maintaining a very small field outside a given circumscribed sphere as sketched in Figure \ref{PhoneNearGeom} where, not appearing in the sketch, the fields are required to decay to very small levels beyond a radius of 1.5 meters. The main difficulty in this application is the fact that we want a good control in the designated listening region with a very fast decay of the fields while maintaining a rather low uniform energy distribution in the region between the sources and the control areas for an overall smooth sound scene.

To address these issues, we propose a scheme that takes advantage of our previous works  \cite{Onofrei2014, Platt2018, Egarguin2018, EgarguinWM2020} and employs a Fourier synthesis scheme towards a time-domain control effect. The central idea of our solution requires field pattern matching in the control regions with fast enough decay of the generated acoustic field so that either the reverberations from the boundary $\partial R$ are insignificant insofar as creating unwanted interference in the sound produced in each control region for the car cabin scenario, or that very little perturbation outside the listening area in the shielded communication application. Details of this scheme are discussed in the following section.

\section{Mathematical Framework }
\label{math}
In this section, we discuss the mathematical framework of our proposed solution to the acoustic control problem in enclosed environments described in the previous section. The foundation of our proposed solutions rests on the possibility of creating an almost non-radiating acoustic field in the interior of the region $R$ using the sources $D_j$, $j =\overline{1,n}$ with the requirement that this generated field approximates the prescribed fields $f_p$ on the interior control regions $R_p$. The framework described below were established in previous works \cite{Onofrei2014, Platt2018, Egarguin2018, EgarguinWM2020}.

Let $\{R_1, R_2,...,R_m\}$ be a collection of $m$ mutually disjoint smooth domains inside $R \subseteq \mathbb R^3$ where $R_m$ is the exterior of some large enough null sphere $ R \setminus B_r(\B0)$ for some prescribed value of $r>0$. These domains will be referred to as the control regions. The problem  is to generate a prescribed acoustic field in $R_p$, $p= \overline{1,m}$ using a collection of active sources $\{D_1, D_2,..., D_n\}$ modeled as mutually disjoint compact regions in $R$ with Lipschitz continuous boundaries. It is assumed that the sources and control regions are well-separated from each other.  Formally, the problem is to find some appropriate  boundary input on these sources, either a normal velocity $v_j \in C(\partial D_j )$ or an acoustic pressure $p_j \in C(\partial D_j)$, $j = \overline{1,n}$, such that for any desired field $f = (f_1, f_2,..., f_m)$ on the control regions, the solution $u$ of the following problem 
\begin{equation}
\label{P1a}
\vspace{0.15cm}\left\{\vspace{0.15cm}\begin{array}{llll}
\GD u+k^2u=0 \mbox{ in }\RR^3\!\setminus \left ( \displaystyle \bigcup_{j=1}^n \!\overline{D}_j \right ) \vspace{0.15cm},\\
\nabla u\cdot \Bn_{j}=v_j, (\mbox{ or } u=p_j)\mbox{ on }\partial D_j\vspace{0.15cm}, ~j= \overline{1,n}\\
\displaystyle\left<{\hat{x}},\nabla u(x)\right>\! -\!iku(x)\!=\!o\left(\frac{1}{|x|}\!\right)\!,\mbox{ as }|x|\rightarrow\infty
\mbox{ uniformly for all ${\hat{x}}$, }\end{array}\right.
\end{equation}
satisfies the control constraint \begin{equation} \label{P1b} \Vert u-f_p\Vert_{C^2(R_p)}\leq \mu \text{ for } l= \overline{1,m}
 \end{equation}  where $\mu$ is some small positive parameter, $C^2(M)$ denotes the space of functions defined on $M$ with continuous partial derivatives up to the second order, and $\Bn_{j}$ is the outward unit normal to $\partial D_j$ and ${\hat{x}}=\frac{x}{|x|}$ denotes the unit vector along the direction $x$. As a convention, the $e^{-i \omega t}$ ($w=kc$) of the fields is implicitly assumed and omitted for the simplicity of analysis in the fixed frequency case. 

Following the formulation in  \cite{Onofrei2014}, the solution to \eqref{P1a} - \eqref{P1b} is assumed to be a layer potential. Provided $k$ is not a resonance (see \cite{colton_kress} and \cite{Onofrei2014}, \cite{Platt2018}), the boundary input (normal velocity $v_j$ or pressure $p_j$) on each source $D_j$ can be characterized by a density function $w_j$ such that
\begin{eqnarray}
& v_j(x) =&\displaystyle \frac{-i}{\rho c k}\frac{\partial}{\partial\Bn_{1}}\int_{\partial D'_{j}}w_j(y) \phi_\omega(x,y)~dS \text{ and }\label{eqnvn}\\
\nonumber \\ 
& p_j(x)=&\displaystyle \int_{\partial D'_{j}}w_j(y)	\phi_\omega(x,y) ~dS,\label{eqnpb}
\end{eqnarray} where $D'_j$ is a fictitious source that is compactly embedded in $D_j$ and $\phi_\omega$ is the fundamental solution of the Helmholtz problem at fixed frequency $\omega=kc$ in free space. Although the expressions in \eqref{eqnvn} and \eqref{eqnpb} make use of the single layer potential operator,  it was shown in \cite{Platt2018} (see also \cite{Egarguin2018}) that appropriate boundary inputs can be obtained by using the single layer potential representation or more generally, a representation given by a linear combination of single and double layer potentials which can actually remove the condition that $k$ is not a resonance. Moreover, it was noted in the aforementioned works that the control of the acoustic fields in the entire volume of each control region $R_p$ can be ensured by controlling the acoustic field on the surface of  a slightly larger control region $W_p$ that compactly embeds $R_p$, by appealing to the uniqueness and regularity properties of the solutions of the interior problem \eqref{P1a}.

In \cite{EgarguinWM2020}, the solution of the forward interior problem \eqref{P1a} was defined through the operator \[\calD:  \displaystyle \prod_{j=1}^n L^2(\partial D'_j) \to \displaystyle \prod_{p=1}^m L^2(\partial W_p)\]
that takes as input the density function $w = (w_1, w_2, ...w_n)$, whose components are defined on the surface of the fictitious sources, that characterize the boundary input via \eqref{eqnvn} or \eqref{eqnpb} and produces the output $\calD w$ representing the generated field on the surface of the slightly larger control regions $(W_1, W_2, ..., W_m)$. It should be noted that the operator $\calD$ takes into account the first-order coupling among the sources. The two main results of \cite{EgarguinWM2020} pertinent to this study are as follows.

\begin{thm}{[\cite{EgarguinWM2020}, Theorem 2.2]}
\label{existence_result}
The compact operator $\calD$  has a dense range.
\end{thm}

\begin{thm}{[\cite{EgarguinWM2020}, Proposition 3.1]}
\label{stability_result}
The solution scheme proposed in \cite{EgarguinWM2020} is stable with respect to perturbations of the prescribed field to be synthesized.
\end{thm}

Theorem \ref{existence_result} guarantees that for any Helmholtz field $f = (f_1, f_2,..., f_m) \in \displaystyle \prod_{p=1}^m L^2(\partial W_p)$, there exists an infinity of solutions $w = (w_1, w_2,...,w_n) \in  \displaystyle \prod_{j=1}^n L^2(\partial D'_j)$ such that 
\begin{equation}
\calD w \approx f 
\label{operatorialEqn1}
\end{equation} 
for any given accuracy threshold. Meanwhile, Theorem \ref{stability_result} states that the solution of the perturbed equation 
\begin{equation}
\calD w \approx f + \epsilon s 
\label{operatorialEqn2}
\end{equation} 
where $s$ is a unit noise vector and $0<\epsilon \ll 1$, is close to the solution of the unperturbed equation \eqref{operatorialEqn1}.

\begin{figure}[!h]
\centering
\includegraphics[width=0.5\textwidth]{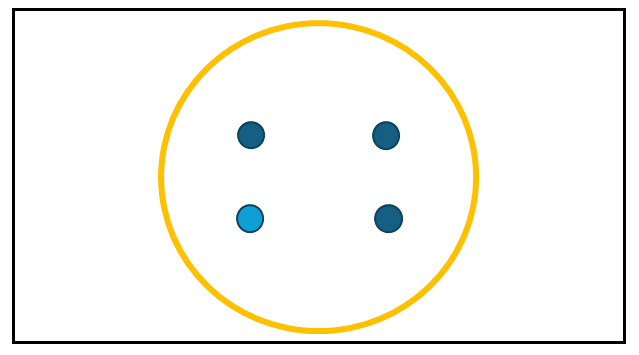}
\caption{An abstraction of the car cabin environment, the solid rectangle represents the enclosure of the cabin while the dots represent the control regions and the orange sphere represents the null sphere}
\label{CarCabin2}
\end{figure}

We take advantage of these results to address the acoustic control problem \eqref{helmholtz1} in the case of enclosed environments with complex boundaries. For instance, in the problem of creating personalized sound zones in car cabins, we consider the abstracted problem geometry shown in Figure \ref{CarCabin2}. The control regions were represented by the blue dots while the boundary of the cabin is illustrated as the solid rectangle. Instead of modeling the boundary condition, we introduce a fictitious null sphere, shown as the orange circle, that encloses the control region but is separated from the boundary. We shall require our acoustic sources to not only approximate the prescribed fields on the control regions, but also to create a near zero acoustic field on the surface of the null sphere. The uniqueness and stability of the exterior problem dictates that the acoustic field beyond the null sphere will be small enough so that the reverberations caused by the reflection of the acoustic fields on the boundary will not have a large impact on the established acoustic profiles on the control regions.

This novel approach removes the necessity of  modeling the complex boundary condition for \eqref{helmholtz1} which may then require tedious calculations and analysis of the appropriate Green's function. In the following sections, we shall present some numerical simulations that illustrates this control strategy.

\section{Numerical Framework}
\label{numerical}

In this section, we discuss the  framework for the numerical solution of  \eqref{P1a}-\eqref{P1b} adapted to the geometry suggested in the previous section where a fictitious null sphere that contains the control regions and is well-separated from the environment's boundary was introduced. The framework was originally broached in the works \cite{Onofrei2014, Platt2018, Egarguin2018, EgarguinWM2020}.

As in\cite{EgarguinWM2020}, we shall use local basis functions to reduce the operatorial equation $\calD w = f$ to a linear system $Ac = b$.  A detailed discussion of local basis functions applied to finite element methods in acoustics and electromagnetics can be found in \cite{Polycarpou2006}. We discretize the surface of each fictitious source $D'_j$ into a set $\{T^1_j,T^2_j, ..., T^N_j\}$  of triangular elements such that each vertex of the triangle lies on the surface of the fictitious source. Then for each vertex $v_i$ of $q^{\text{th}}$ triangle $T^q_j$, we associate a linear interpolating function $N_i$ such that $N_i(v_i) = 1$ and zero on the other two vertices. Thus, for every $ x \in T^q_j$, the unknown density function $w_j$ can be expressed by the linear interpolant
\begin{equation}
\label{w_pieces}
	w_{j, q} (x) = \sum_{i=1}^3 c_{i} N_i(x),
\end{equation}
 where the coefficient $c_{i}$ is the value of $w_j$ on vertex $v_i$. At this point, the continuous problem of finding the density function $w =( w_1, w_2,..., w_n) \in  \displaystyle \prod_{j=1}^n L^2(\partial D'_j)$ is reduced to solving for the coefficients $c_{i} \in \mathbb C$.

To fully discretize the equation $\calD w = f$, we shall use collocation test points from a dense mesh of points from each $W_p$. If  there is a total of $M$ discretization points on the control regions and a total of $N$ points on the surface of the fictitious sources then the matrices $A$, $c$ and $b$ in the discrete approximation of the operatorial equation are of dimensions $M \times N$, $N \times 1$ and $M \times 1$, respectively.  Each entry of the matrix $A$ is an integral approximated using the standard seven-point Gaussian quadrature rule. Then the unknown vector $c$ of basis coefficients is computed as the Tikhonov solution 
\begin{equation}
\label{c_soln}
c = (\alpha I + A^*A)^{-1}A^*b
\end{equation}
with the regularization parameter  $\alpha$ chosen using the Morozov discrepancy principle. See \cite{Platt2018} for a detailed discussion of the Tikhonov regularization routine using the Morozov discrepancy principle.

The approximate solution for the operatorial equation $\calD w \approx f$ obtained by implementing the mathematical framework described in the previous section through the numerical scheme discussed above appears to be stable with respect to process errors. That is, for a solution $w$ of  $\calD w \approx f$, a unit noise vector $s$ and small enough noise level $\epsilon$ with $0 < \epsilon \ll 1$,  \begin{equation}
\calD (w+\epsilon s)\approx f 
\label{operatorialEqn2}
\end{equation} 
still holds. This can be observed in the simulations in \cite{EgarguinAA2021}, where a detailed feasibility study was conducted for sources in free space and in shallow water environments. These suggest that there is a level of error or noise in the manufacturing of the sources that will not significantly ruin the control effect.

In the next section, we present several simulations illustrating the scheme discussed above. A total of 234 basis functions were used on each spherical fictitious source modeling the actual physical source, while the control regions are discretized depending on the target application. As a computational stability check, these evaluation points are chosen to be the midpoint between consecutive points in the original mesh of collocation points used together with the local basis discretization to obtain the solution \eqref{c_soln} to the associated linear system. We also calculate the pointwise relative error $\dfrac{|f_l(x)-u_l(x)|}{|f_l(x)|}$  and take note of its supremum. Whenever  $W_p$ is the null sphere, we observe the supremum magnitude of the generated field in $W_p$.

\section{Numerical Simulations}
\label{simulations}

In this section, we present the two applications of this work: personal sound zones inside a bounded region and shielded localized communication. The sections present several numerical simulations associated with each of these two applications and illustrate the quality of our controls. 

\subsection{Personal Sound Zones in Car Cabins}
For the case of creating personal sound zones within a car cabin, we shall present two simulations. In the first, we present a static case where we create four planewaves of different frequencies while in the second, we present a time-domain simulation using Fourier synthesis. In both cases, we use a total of eight sources with two sources near each control region. They are all spheres of radius 5 cm positioned relative to their proximal control region in the same way as Figure \ref{CarCabin}. The control regions are also spheres of radius 10 cm positioned inside the car cabin as shown in Figure \ref{CarCabin2}. These control regions, including the null sphere, is discretized into 1250 collocation points. 

\subsubsection{Multi-frequency Simulation}
\label{multifreq}
In this test, we prescribe planewaves  $f_p(\Bx) = e^{i \Bx \cdot (k_p \hat \Bd_p)}, p = \overline{1,4}$ of different propagation directions and frequencies on each control region. The directions of propagation are  $ \Bd_1 = \left < 0, 1, 0 \right >$, $ \Bd_2 = \left < 1, 0, 0 \right >$, $ \Bd_3 = \left < 0, 0, 1 \right >$, and $ \Bd_4 = \left < 0, \frac{\sqrt 2}{2}, \frac{\sqrt 2}{2} \right >$ while the wave numbers are $k_1 = 10, k_2 =15, k_3 = 20, k_p = 25$, corresponding to frequencies 545.90 Hz, 818.85 Hz, 1.09 KHz, and 1.36 KHz, respectively.

\begin{figure}[!h]
\centering
\includegraphics[width=\textwidth]{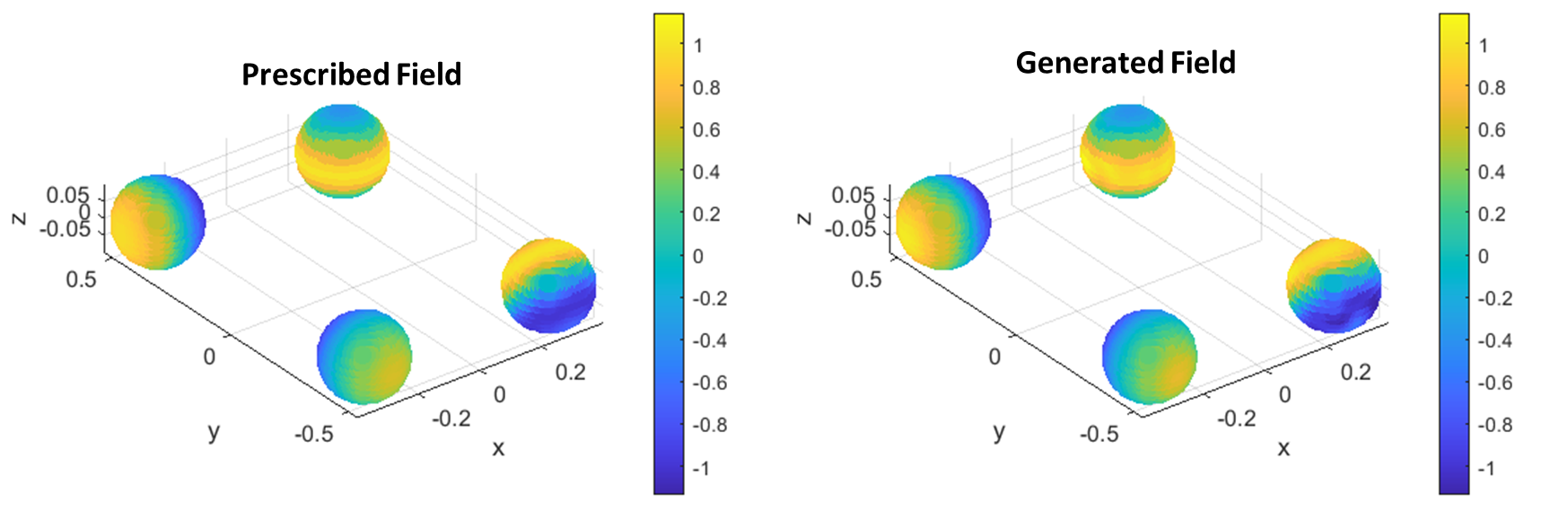}
\caption{A visual comparison of the real part of the prescribed and generated fields}
\label{Sim2Synth}
\end{figure}

\begin{figure}[!h]
\centering
\includegraphics[width=0.5 \textwidth]{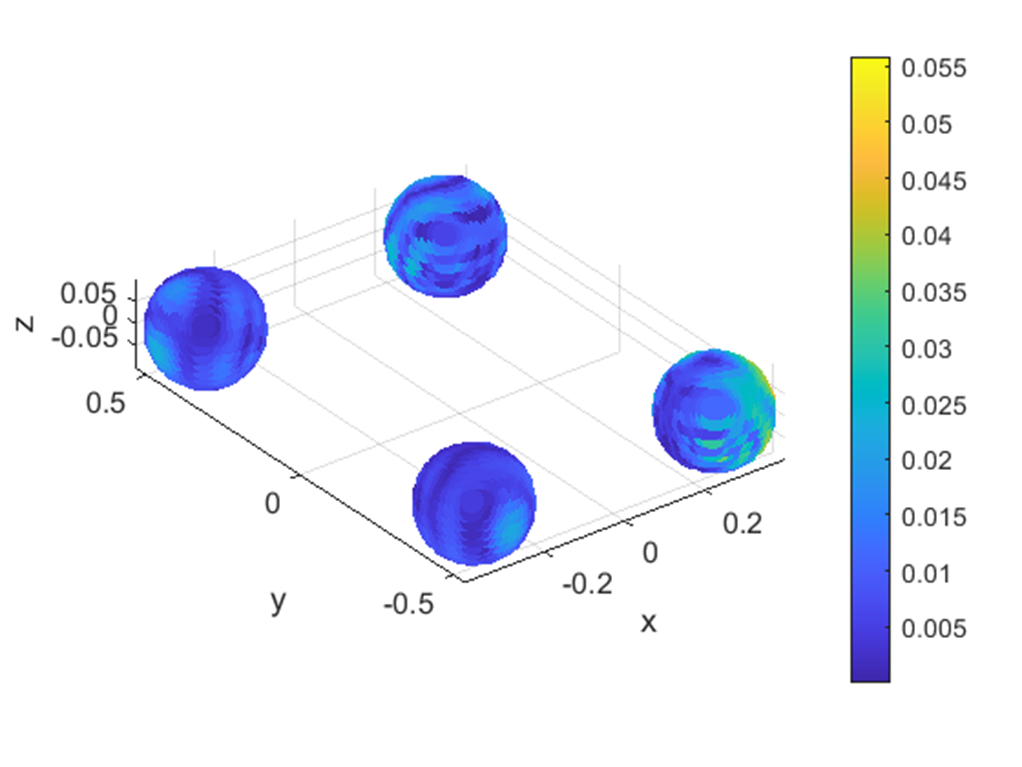}
\caption{Plot of the pointwise relative error}
\label{Sim2relerr}
\end{figure}

Figure \ref{Sim2Synth} shows a very good approximation of the desired patterns in each of the control regions. In fact, the pointwise relative errors are kept below 5.58 \% as seen in Figure \ref{Sim2relerr}.

\begin{figure}[!h]
\centering
\includegraphics[width=0.5 \textwidth]{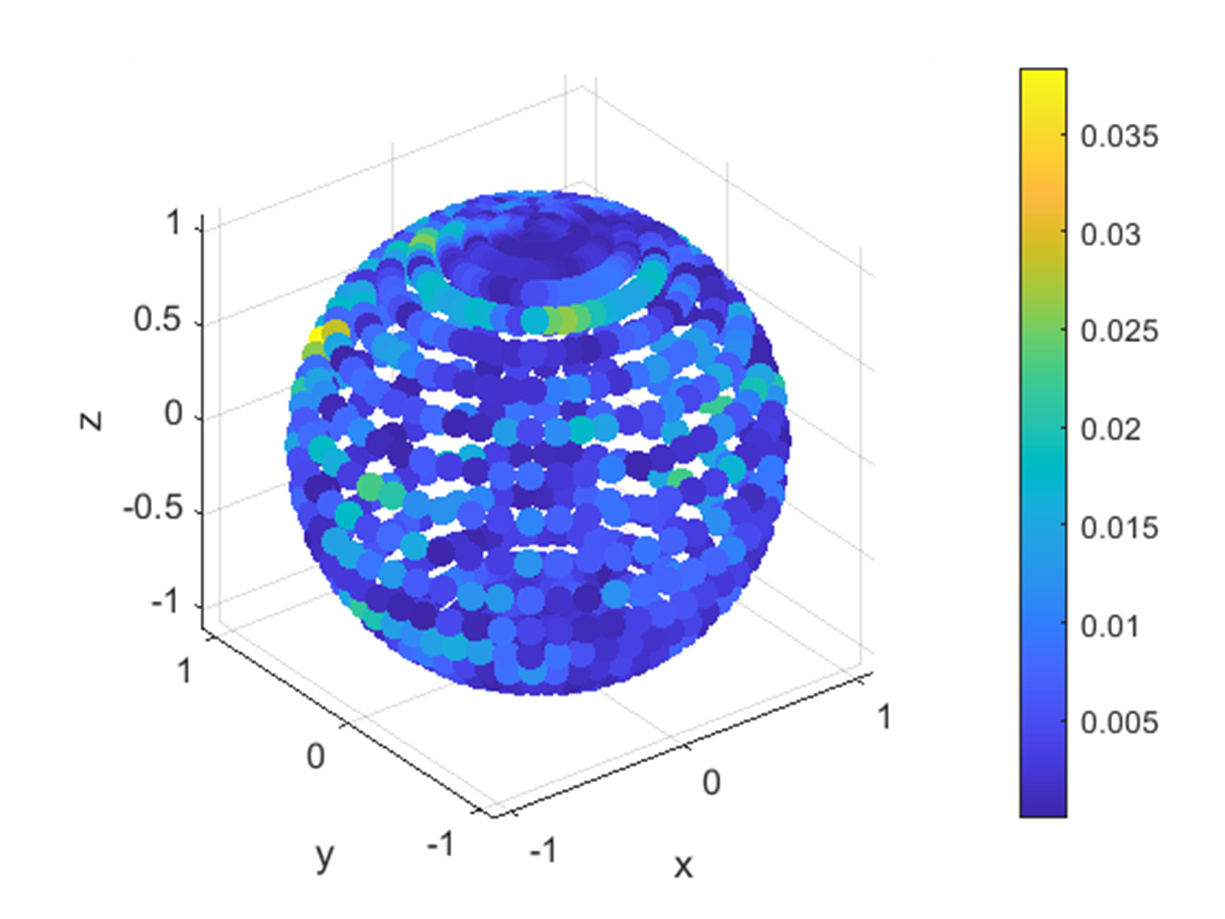}
\caption{Plot of the pointwise magnitude of the generated field in the surface of the null sphere}
\label{Sim2Null}
\end{figure}

Likewise, good results were obtained on the surface of the null sphere. Figure \ref{Sim2Null} shows that the generated field has supremum norm of around $3.88 \times 10^{-2}$ and has values of order $10^{-3}$ in most of the sphere's surface. 

\begin{figure}[!h]
\centering
\includegraphics[width=\textwidth]{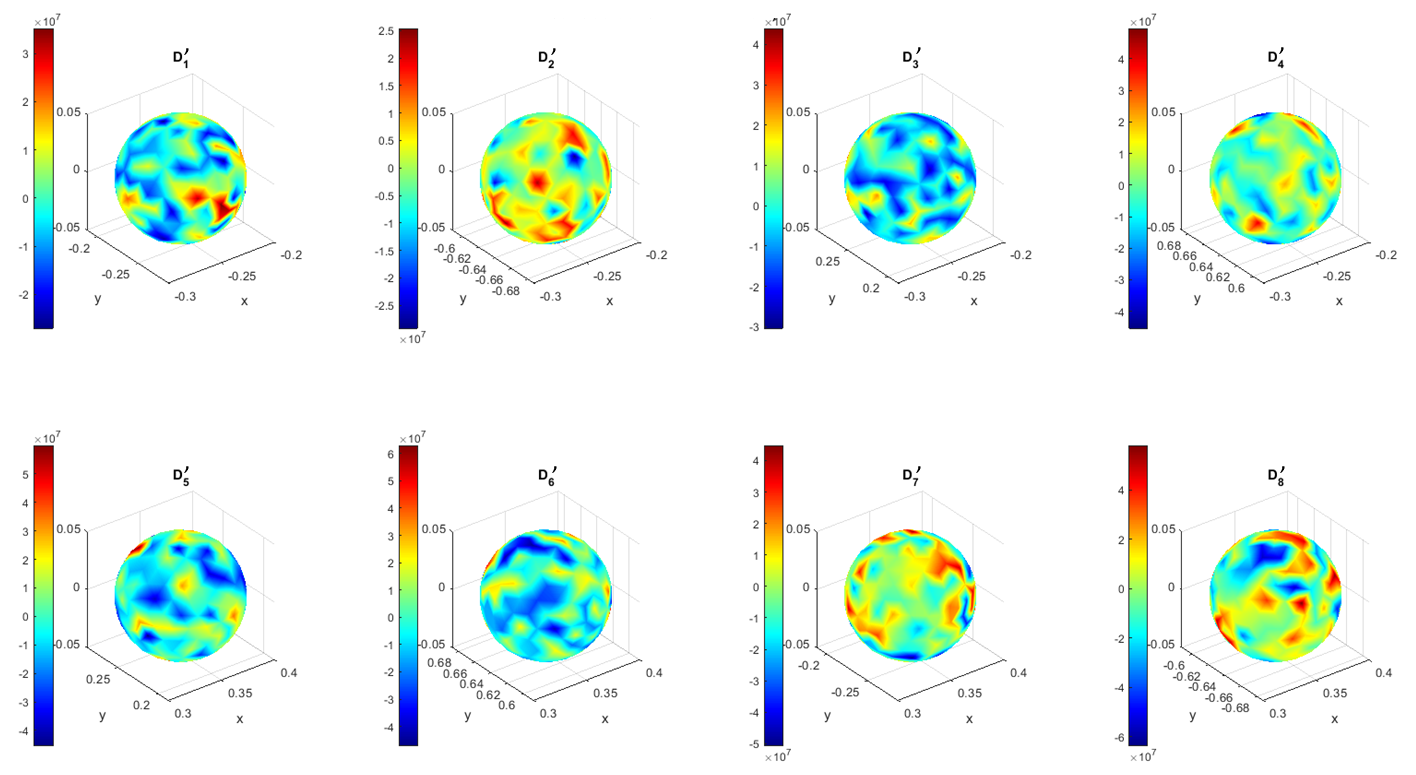}
\caption{Real part of the computed density on the fictitious sources}
\label{Sim2Density}
\end{figure}

\begin{figure}[!h]
\centering
\includegraphics[width= \textwidth]{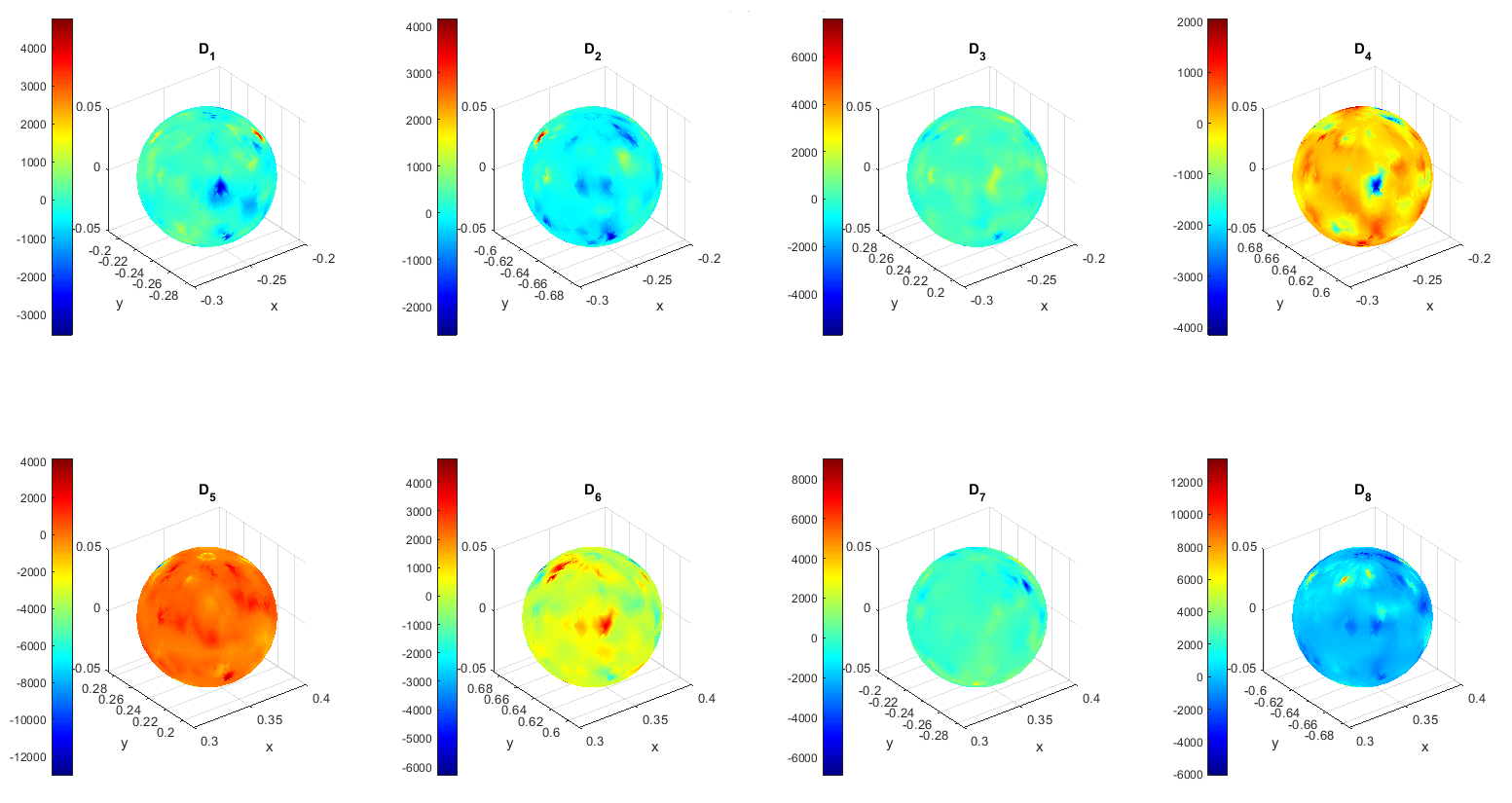}
\caption{Real part of the corresponding normal velocity on the physical sources}
\label{Sim2NV}
\end{figure}

The computed densities on the surface of the fictitious sources $D_j'$ obtained from the numerical scheme and the resulting normal velocities on each physical source $D_j$ are shown in Figure \ref{Sim2Density} and Figure \ref{Sim2NV}, respectively.These show the existence of complex patterns with oscillations that may require non-standard approaches in the physical synthesis of the sources.

This simulation illustrates the possibility of simultaneously creating different sound profiles, even with varying frequency and direction of propagation, on the control regions that will decay fast in space. This fast decay ensures that the reverberations from the boundary of the cabin will not significantly affect the quality of the synthesized field.

\subsubsection{Time-domain Framework}
\label{TimeDomain}

In this subsection, we present the Fourier synthesis scheme for a time-domain extension of the control problem discussed above, a scheme that will again be employed further in the paper for the second time-domain simulation discussing shielded localized sound (presented later in Section \ref{Phone}). 
Assume that the desired target to be approximated is given in the time domain by
\begin{equation}
	z(x, t) = \sum_{j = 1}^N e^{-i \omega_j t} f_j(x) \text{ in } \Omega,
	\label{TD1}
\end{equation}
where $\displaystyle \Omega = \left(\bigcup_{s =1}^{m-1} R_s\right) \bigcup\partial R_m$ is the control region and the $f_j$'s are the solutions of the Helmholtz equation in the neighborhoods of $R_s, s = \overline{1, m}$ with frequency $\omega_j$ for $j = \overline{1, N}$. We will approximate the field in \eqref{TD1} by building time harmonic approximations for each individual frequency and then employing superposition through Fourier synthesis.

Let $\displaystyle \bigcup_{j=1}^n D'_j = D'_a \subset \bigcup_{j=1}^n D_j = D_a$ and $\omega = \omega_0$ be a fixed frequency. We want to first find the time domain source input $q(y, t)$ which will produce the desired radiation to be approximated $e^{-i \omega_0 t}f(x)$ where $f$ solves the Helmholtz equation in a neighborhood of $\Omega$. Let
\begin{equation}
	q(y, t) = e^{-i \omega_0 t} w(y) \big ( H(t) - H(t-T) \big )
	\label{TD2}
\end{equation}
where $H(t) = \begin{cases} 1, & t > 0 \\ 0, &t < 0 \end{cases}$ is the Heaviside step function and $w$ is a source density distributed on $\partial D'_a = \displaystyle \bigcup_{j = 1}^n \partial D'_j$ and $T> 0$ is the large enough final time to be chosen later. Then by Fourier transform, we have that the acoustic field corresponding to this time domain source is given by
\begin{equation}
	v(x, t) = \dfrac{1}{2\pi} \int_{-\infty}^{+\infty} \int_{\partial D'_a} \phi_\omega(x, y) \hat{q}(y, \omega)e^{-i \omega t} dS_y ~d\omega
	\label{TD3}
\end{equation}
where 
\begin{equation}
	\hat{q} (y, \omega) = \int_{-\infty}^{+\infty} q(y, t) e^{i \omega t} dt = \int_{0}^{T} q(y, t) e^{i \omega t} dt
	\label{TD4}
\end{equation}
based on \eqref{TD2}. Using \eqref{TD2} in \eqref{TD4} yields
\begin{align}
	\hat{q}(y, \omega) &= \int_{-\infty}^{\infty} e^{-i \omega_0 t} w(y) e^{i \omega t} \big ( H(t) - H(t-T) \big ) \nonumber \\
	& = w(y) \left [ \int_{-\infty}^{+\infty} e^{i (\omega - \omega_0)t} H(t) dt - \int_{-\infty}^{+\infty} e^{i (\omega - \omega_0)} H(t- T) dt  \right ] \nonumber \\
	& =  w(y) \big (\hat H(\omega - \omega_0) -e^{i (\omega - \omega_0)T} \hat H(\omega - \omega_0) \big ), \label{TD5}
\end{align}
where we applied the change of variable $t = \tilde t + T$ in the last integral above and where $\hat H$ denotes the Fourier transform of $H$.

Using \eqref{TD5} in \eqref{TD3}, we obtain
\begin{equation}
	v(x, t) = \dfrac{1}{2 \pi} \int_{\infty}^{+\infty} \int_{\partial D'_a} \phi_\omega(x, y) w(y) \hat H(\omega - \omega_0) \left ((1 - e^{i (\omega - \omega_0) T} \right) e^{-i \omega t} dS_y ~d\omega.
	\label{TD6}
\end{equation}
Note that
\begin{align}
	\phi_\omega(x, y) & = \dfrac{e^{ik |x-y|}}{4 \pi |x-y|} = \dfrac{e^{i \omega \frac{|x-y|}{c}}}{4\pi |x-y|} = \dfrac{e^{i (\omega - \omega_0)\frac{|x-y|}{c}}}{4\pi |x-y|}  \cdot e^{i \omega_0 \frac{|x-y|}{c}} \nonumber \\
	& = \phi_{\omega_0}(x, y) \cdot e^{i (\omega - \omega_0) \frac{|x-y|}{c}}, \label{TD7}
\end{align}
where $c$ is the speed of sound in the medium. Using \eqref{TD7} in \eqref{TD6}, we get
\begin{align}
	v(x, t) & = \dfrac{1}{2\pi} \int_{-\infty}^{+\infty} \int_{\partial D'_a} \phi_{\omega_0}(x, y) w(y) e^{i (\omega - \omega_0) \frac{|x-y|}{c}}\hat H(\omega - \omega_0) \big (1 - e^{i (\omega - \omega_0)T} \big ) e^{-i \omega t} dS_y ~d\omega \nonumber \\
	& =  \dfrac{1}{2\pi} \int_{\partial D'_a} \phi_{\omega_0}(x, y) w(y) \int_{-\infty}^{+\infty}  e^{i (\omega - \omega_0) \frac{|x-y|}{c}}\hat H(\omega - \omega_0) \big (1 - e^{i (\omega - \omega_0)T} \big ) e^{-i \omega t} d\omega ~ dS_y.
	\label{TD8}
\end{align}
Note that 
\begin{align}
	& \int_{-\infty}^{+\infty}  e^{i (\omega - \omega_0) \frac{|x-y|}{c}}\hat H(\omega - \omega_0) \big (1 - e^{i (\omega - \omega_0)T} \big ) e^{-i \omega t} d\omega \nonumber \\
	= & e^{-i \omega_0 t} \int_{-\infty}^{+\infty} e^{i \tilde \omega \frac{|x-y|}{c}} \hat H(\tilde \omega) \big (1-e^{i \tilde \omega T} \big ) e^{-i \tilde \omega t} d \tilde \omega \nonumber \\
	= & e^{-i \omega_0 t} \cdot 2\pi \cdot \left [ H \left ( t - \dfrac{|x-y|}{x} \right )- H \left ( t -T - \dfrac{|x-y|}{c}\right ) \right ] \label{TD9}
\end{align}
where we used the change of variable $\omega - \omega_0 = \tilde \omega$. Using \eqref{TD9} in \eqref{TD8} gives
\begin{equation}
	v(x, t) = e^{-i \omega_0 t} \int_{\partial D'_a} \left [ H \left (t- \dfrac{|x-y|}{c}\right ) - H \left (t- T - \dfrac{|x-y|}{c}\right )\right ] \phi_{\omega_0}(x, y) w(y) dS_y.
\end{equation}
Thus, if the density $w$ is such that 
\begin{equation}
	\int_{\partial D'_a} \phi_{\omega_0}(x, y) w(y) dS_y \approx f(x) \text{ in } \Omega.
	\label{TD11}
\end{equation}
From \eqref{TD11}, we will have that for given $x \in \Omega$
\begin{equation}
	v(x, t) \approx e^{-i \omega_0 t} f(x) ,
	\label{TD12}
\end{equation}
 in the time interval prescribed by
\begin{equation}
	\max_{y \in \partial D_a} \dfrac{|x-y|}{c} < t < \min_{y \in \partial D_a} \dfrac{|x-y|}{c}+ T.
	\label{TD13}
\end{equation}
Thus, from \eqref{TD12} and \eqref{TD13} we obtain that \[ v(x, t) \approx e^{-i \omega_0 t} f(x) \mbox{ holds for all $x \in \Omega$}, \]  in the following time interval:
\begin{equation}
	t_1=\max_{x \in \Omega} \max_{y \in \partial D_a} \dfrac{|x-y|}{c} < t < \min_{x \in \Omega} \min_{y \in \partial D_a} \dfrac{|x-y|}{c}+ T=t_2.
	\label{TD14}
\end{equation}
Note that \eqref{TD14} imposes a restriction on $T$ and the geometry of our problem as, for example, \eqref{TD14} necessarily implies that 
\begin{equation}
	t_1=\max_{x \in \Omega} \max_{y \in \partial D_a} \dfrac{|x-y|}{c} <  \min_{x \in \Omega} \min_{y \in \partial D_a} \dfrac{|x-y|}{c}+ T=t_2.
	\label{TD15}
\end{equation}

By superposition and repeating the above analysis for every $\omega = \omega_j$ in \eqref{TD1} we have that the field radiated by the time domain source density $q \in C \big (\partial D_a',  [0, T] \big )$ given by 
\begin{equation}
	q(y, t) = \sum_{j =1 }^N e^{-i \omega_j t} w_j(y) \big ( H(t) - H(t -T) \big )
	\label{TD16}
\end{equation}
when $w_j \in C(\partial D'_a)$ is such that
\begin{equation}
	\int_{\partial D'_a} \phi_{\omega_j}(x, y) w_j(y) dS_y \approx f_j(x) \text{ in } \Omega.
	\label{TD17}
\end{equation}
Here, the approximation is of order $\dfrac{\mu}{N}$ in $C^2(\Omega)$ where $\mu$ was the desired accuracy level introduced in \eqref{P1b}. Hence, 
\begin{equation}
	||v-\sum_{j =1}^N e^{-i \omega_j t}f_j||_{(C[t_1,t_2],C^2(\Omega))}<\mu
	\label{TD18}
\end{equation}
This implies that the desired control effect \eqref{TD18} will not be felt in the time interval immediately after the opening of the sources, i.e., for $t \in \left ( 0, \displaystyle \max_{x \in \Omega} \max_{y \in \partial D_a} \dfrac{|x-y|}{c} \right )$. Then, the effect will be felt a bit after the sources shut down, i.e.,  in the time interval $ \left ( T, \displaystyle \min_{x \in \Omega} \min_{y \in \partial D_a} \dfrac{|x-y|}{c}+T \right )$. We can further implement delays and different stop times for each source so that we minimize the interval of incoherence and make the control effect smoother.

\subsubsection{Time-domain Simulation}
\label{TimeDomainSimulation}

In what follows, we illustrate the above time-domain discussion in a simulation of the control problem inside a car cabin. We adopt the same geometrical configuration as in Section \ref{multifreq}. We run a multitude of test in each of which we impose the planewave $f_q(\Bx) = e^{i \Bx \cdot (k_q  \left < 0, 1, 0 \right >)}$ on $D_1$ (the lower left control region in Figure \ref{CarCabin2}) with $k_q = \overline {1, 50}$ and zero fields on the remaining three control regions, as well as the null sphere. The time-dynamic prescribed field $f$ and generated field $u$ are then obtained by superimposing the fields  from the different frequencies scaled by the time dependent factor $e^{-ik_qct}$, i.e., 

\[f(\Bx, t) = \sum_{q=1}^{50} f_q (\Bx) e^{-ik_qct} \text{ and } u(\Bx, t) = \sum_{q=1}^{50} u_q (\Bx) e^{-ik_qct} \] where $c = 343$ m/s is the assumed speed of sound in the medium. We let $t \in [0, 0.04]$ so that the time-domain animations will show around two periods of the generated waves.

\begin{figure}[!h]
\centering
\includegraphics[width= 0.9 \textwidth]{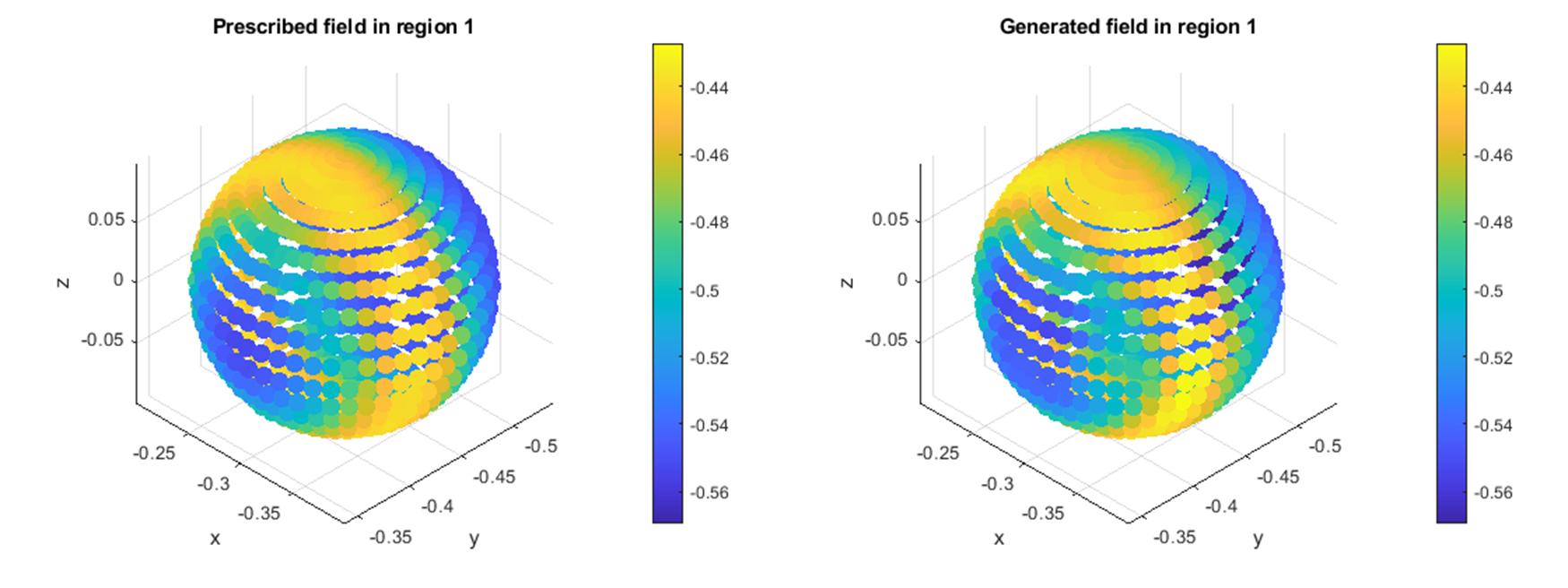}
\caption{A visual comparison of the time-averaged real part of the prescribed and generated fields on $D_1$}
\label{Sim3SynthPlane}
\end{figure}

\begin{figure}[!h]
\centering
\includegraphics[width=0.5 \textwidth]{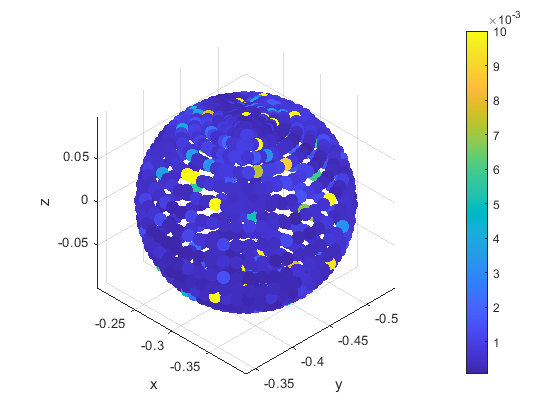}
\caption{Plot of the time-averaged pointwise relative error}
\label{Sim3relerr}
\end{figure}

The results of the time-domain synthesis are shown in an animation found in \cite{Sim3NearControl}. The time-averaged plots of the prescribed and generated fields are shown in Figure \ref{Sim3SynthPlane}. This shows a good visual match between the fields. The time-averaged relative error is around around 0.85 \%. The plot of the pointwise relative errors is shown in Figure \ref{Sim3relerr} while the time evolution of the pointwise relative error can be found in this animation \cite{Sim3RelErr}. 

\begin{figure}[!h]
\centering
\includegraphics[width=0.6\textwidth]{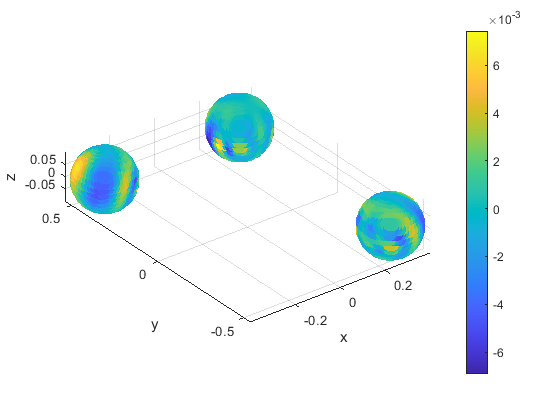}
\caption{Plot of the time-averaged real part of the generated field on the remaining three control regions}
\label{Sim3SynthNull}
\end{figure}

The results on the other three near control regions are likewise good. The time-domain evolution is shown in the animation \cite{Sim3NullControlsAnimation}, while the time-averaged field values are shown in Figure \ref{Sim3SynthNull}. The figure shows that on the average, the field values are of order $10^{-3}$, way below the magnitude of the field generated on $D_1$.

\begin{figure}[!h]
\centering
\includegraphics[width=0.5 \textwidth]{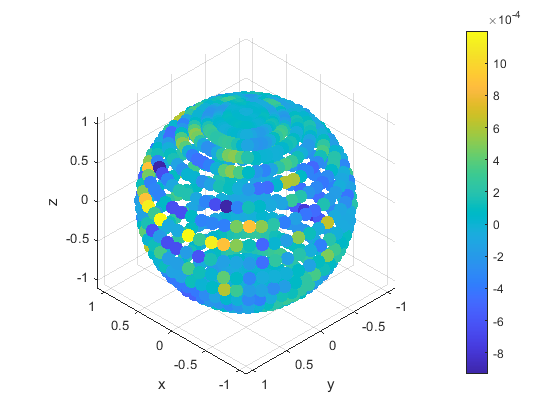}
\caption{Plot of the time-averaged real part of the generated field in the surface of the null sphere}
\label{Sim3Null}
\end{figure}

Lastly, the field synthesized on the surface of the  null sphere is also of small amplitude as seen in the time evolution animation found in \cite{Sim3NullRegionAnimation} and in the time-averaged plot in Figure \ref{Sim3Null}. The supremum of the time-averaged field is around $1.4 \times 10^{-3}$.

We also mention here that in order to recreate personal sound experiences in each of the 4 control regions we can repeat the above procedure for each of the 4 regions and then superimpose the results to obtain the desired fast decaying field with controlled patterns in each of the four control regions. Also as observed in the analysis performed in Section \ref{TimeDomain} there will be an initial time interval $[0,t_1]$ with $t_1$ defined at \eqref{TD14}, during which the desired control effect will not be observed. This interval is very small considering that the speed of sound is much larger than the distances considered in our problems and can further be minimized by introducing appropriate delays at each of the point sources in the overall array. This is why we believe that the effect of this short lived disturbance will be small when interacting with the boundary of $R$. 

\begin{figure}[!h]
\centering
\includegraphics[width=\textwidth]{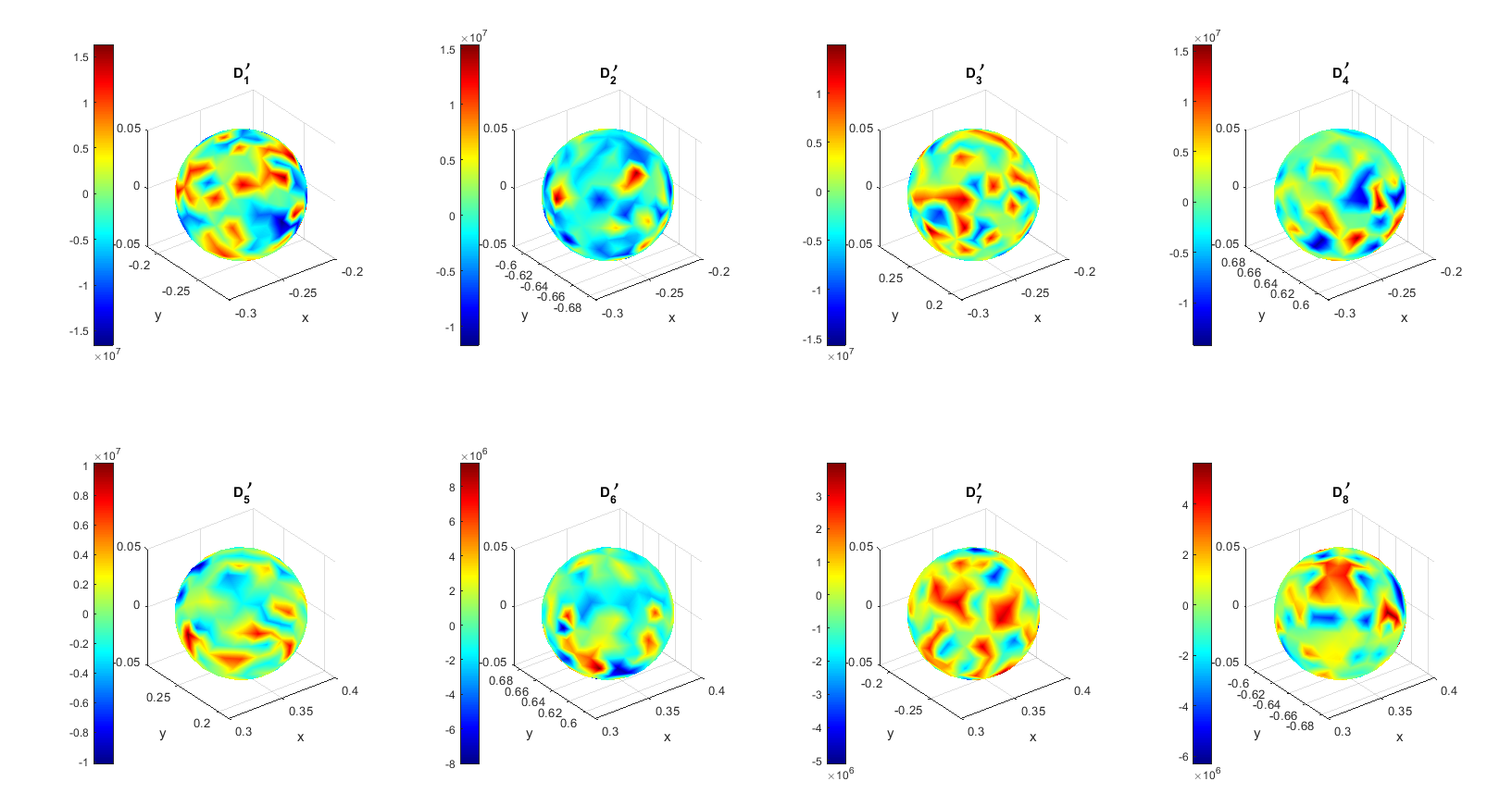}
\caption{Time-averaged real part of the computed density on the fictitious sources}
\label{Sim3Density}
\end{figure}

\begin{figure}[!h]
\centering
\includegraphics[width= \textwidth]{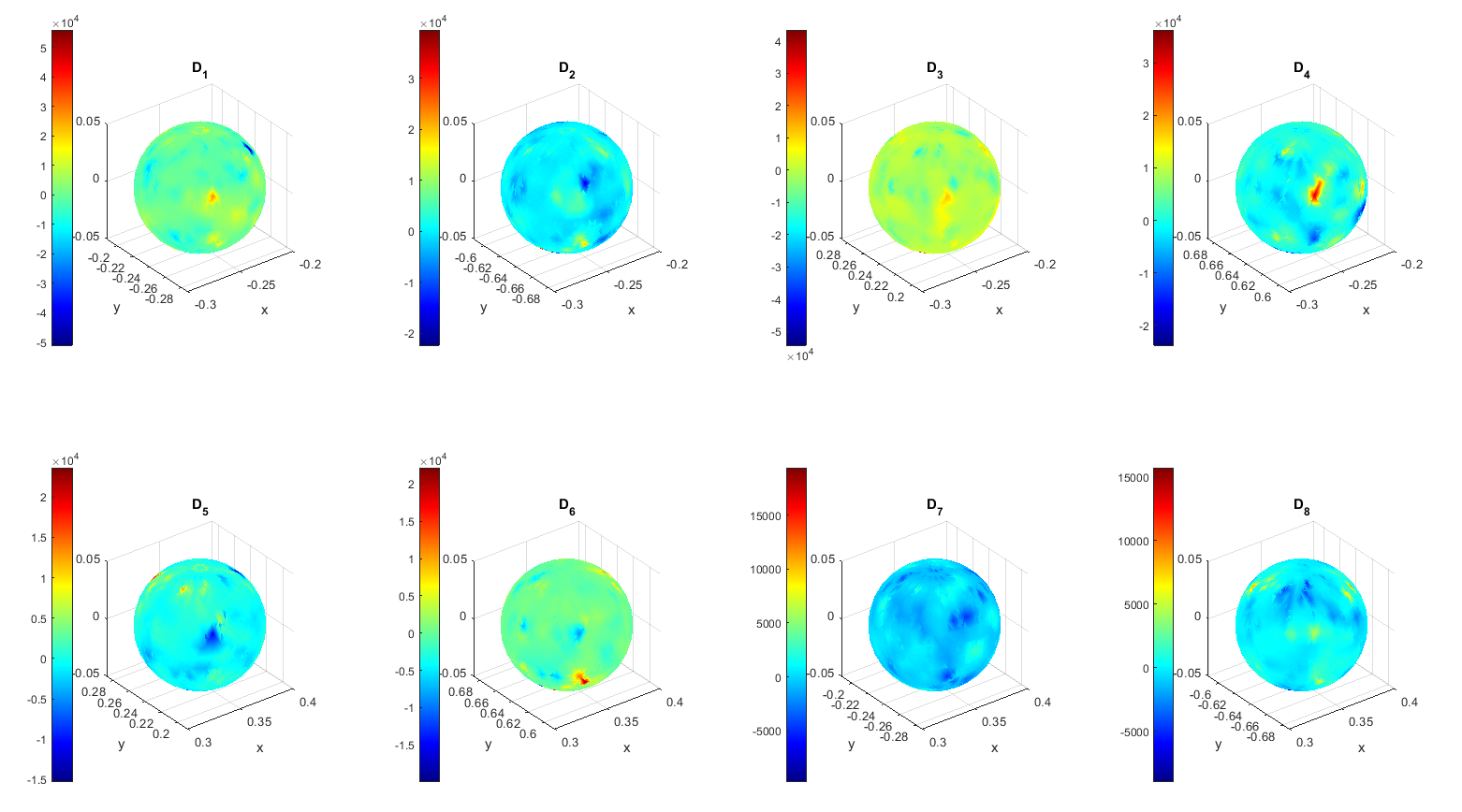}
\caption{Time-averaged real part of the corresponding normal velocity on the physical sources}
\label{Sim3NV}
\end{figure}

The time-averaged real part of the computed density and its corresponding normal velocity on each source are shown in Figure\ref{Sim3Density} and Figure \ref{Sim3NV} while the animated time-evolution can be found in \cite{Sim3DensityAnimate} and \cite{Sim3NVAnimate}, respectively. Even in the time domain, both the density and normal velocity exhibits a highly oscillatory and complex behavior.

\subsection{Shielded Localized Communications}
\label{Phone}
In this simulation, we address the problem of creating shielded localized communication, i.e., establishing a desired sound profile in a given region exterior to the source array with very small field beyond a certain small radius, away from the array. We model the listening zone as a spherical control region of radius 25 cm with center at $(-0.52, 0, 0.20)$. Meanwhile, the source array is modeled by 12 spherical sources $D_i, i= \overline{1, 12}$ with radius 1 cm whose centers are uniformly distributed on the sides of a 7 cm by 15 cm rectangle (usual phone dimensions). The center of the rectangle is assumed to be at the origin, making its center 20 cm below the center of the control region.  See Figure \ref{PhoneNearGeom}. The far control region, outside of which we want the acoustic field to be very small,  is a sphere of radius 1.5 m concentric with the rectangle. 

To demonstrate our scheme, we wish to create the plane wave  $f(\Bx) = e^{i \Bx \cdot (k \hat \Bd_1)}$, where  $ \Bd_1 = \left < 0, 1, 0 \right >$ and wave number $k = 10$ (corresponding to about 546 Hz) on the near control while imposing a null field on the far control. Both control regions were discretized into 3200 collocation points.

\begin{figure}[!h]
	\centering
	\includegraphics[width=0.8 \textwidth]{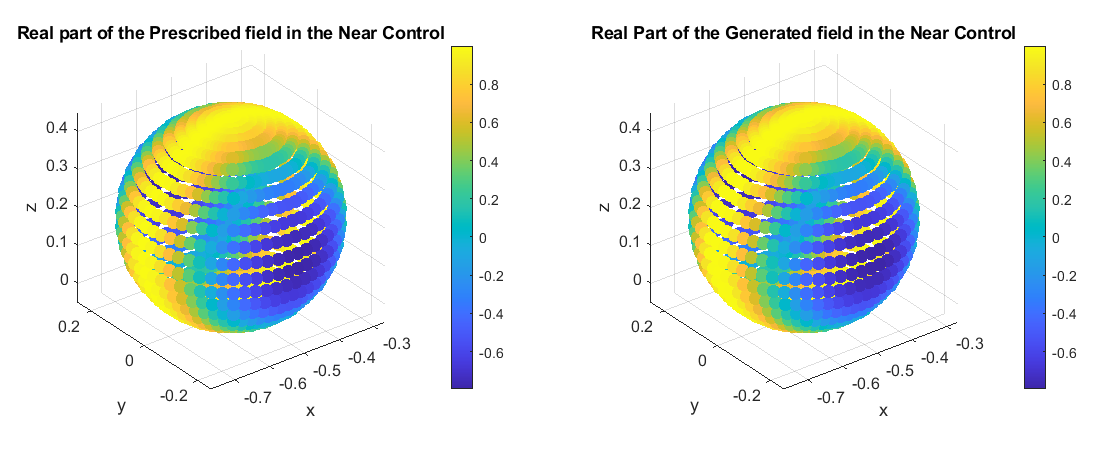}
	\caption{A visual comparison of the real part of the prescribed and generated fields on the control region}
	\label{PhoneSimNearControl}
\end{figure}

\begin{figure}[!h]
	\centering
	\includegraphics[width=0.5 \textwidth]{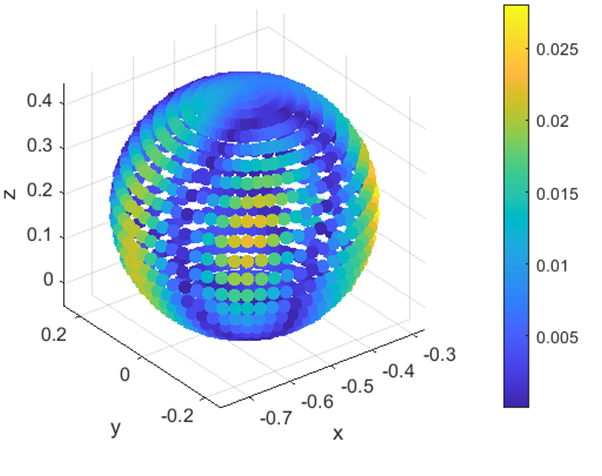}
	\caption{Plot of the pointwise relative error}
	\label{PhoneSimRelErr}
\end{figure}

The results of the field synthesis on the near control regions are shown in Figure \ref{PhoneSimNearControl}. A visual comparison of these fields suggest a good match between the prescribed and generated fields. It is confirmed in Figure \ref{PhoneSimRelErr} where we see the pointwise relative errors. We see that throughout the near control, the relative error is just below 3\%.

\begin{figure}[!h]
	\centering
	\includegraphics[width=0.5 \textwidth]{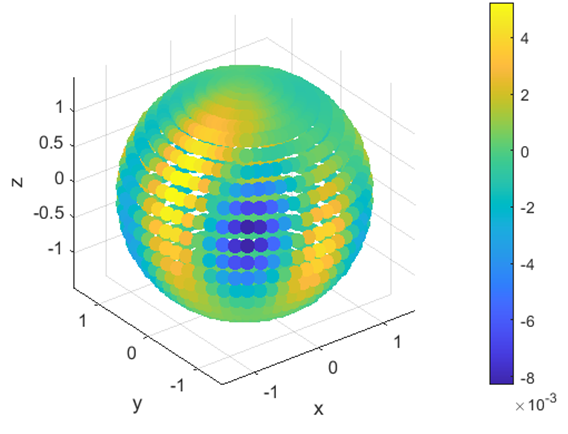}
	\caption{Plot of the real part of the generated field in the surface of the null sphere}
	\label{PhoneSimFarControl}
\end{figure}

Meanwhile, the generated field on the surface of the null sphere is shown in Figure \ref{PhoneSimFarControl}. It can be noted that the generated field has supremum $8.29 \times 10^{-3}$, which is three orders lower than the field on the near control. This guarantees little spill over of the acoustic pressure that will ensure that observers beyond the far control will not receive an intelligible signal and obstacles beyond the sphere will not cause significant echo or reverberations.

\begin{figure}[!h]
	\centering
	\includegraphics[width= \textwidth]{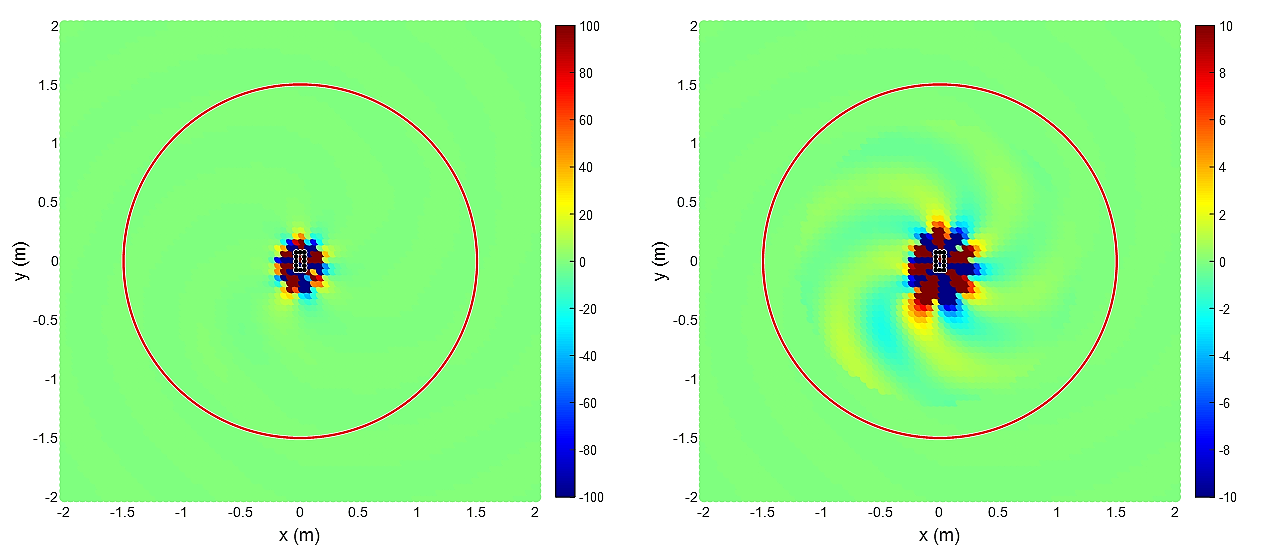}
	\caption{Real part of the generated field on the plane $z=0$,  the color map on the left plot is truncated to $[-100, 100]$ while the right plot's color bar is set to $[-10, 10]$ to show the radiated pattern in different scales}
	\label{PhoneSimCutPlane0p0}
\end{figure}

\begin{figure}[!h]
	\centering
	\includegraphics[width= \textwidth]{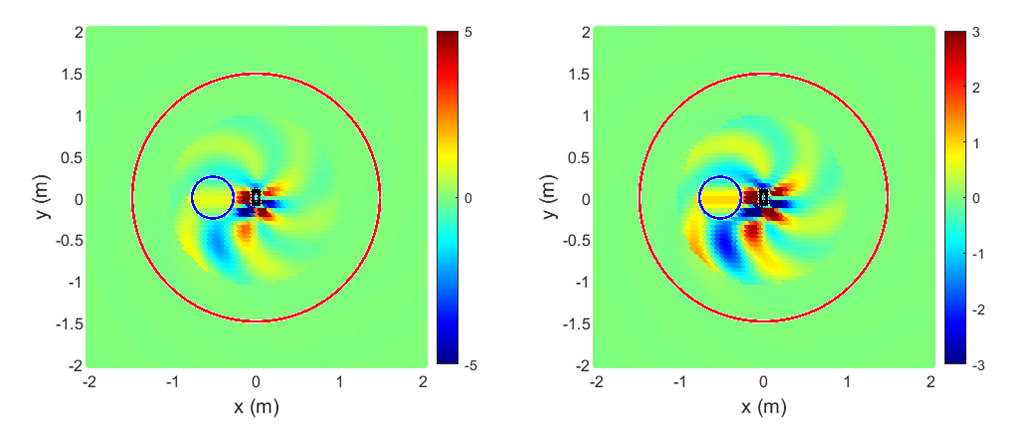}
	\caption{Real part of the generated field on the plane $z=0.2$,  the color map on the left plot is truncated to $[-5, 5]$ while the right plot's color bar is set to $[-3, 3]$ to show the radiated pattern in different scales}
	\label{PhoneSimCutPlane0p0-1}
\end{figure}

Figure \ref{PhoneSimCutPlane0p0} and Figure \ref{PhoneSimCutPlane0p0-1} shows the generated field on the plane $z=0$ and $z=0.2$ respectively. The red circle is the projection of the far control on the planes while the black circles near the origin represent the projection of the boundary of the sources. The color maps on these plots are truncated to better show the variations in the magnitude of the field. These show that most of the radiated energy is focused on the area near the sources while decaying fast enough to achieve near zero magnitude on the far control region.

\begin{figure}[!h]
	\centering
	\includegraphics[width=\textwidth]{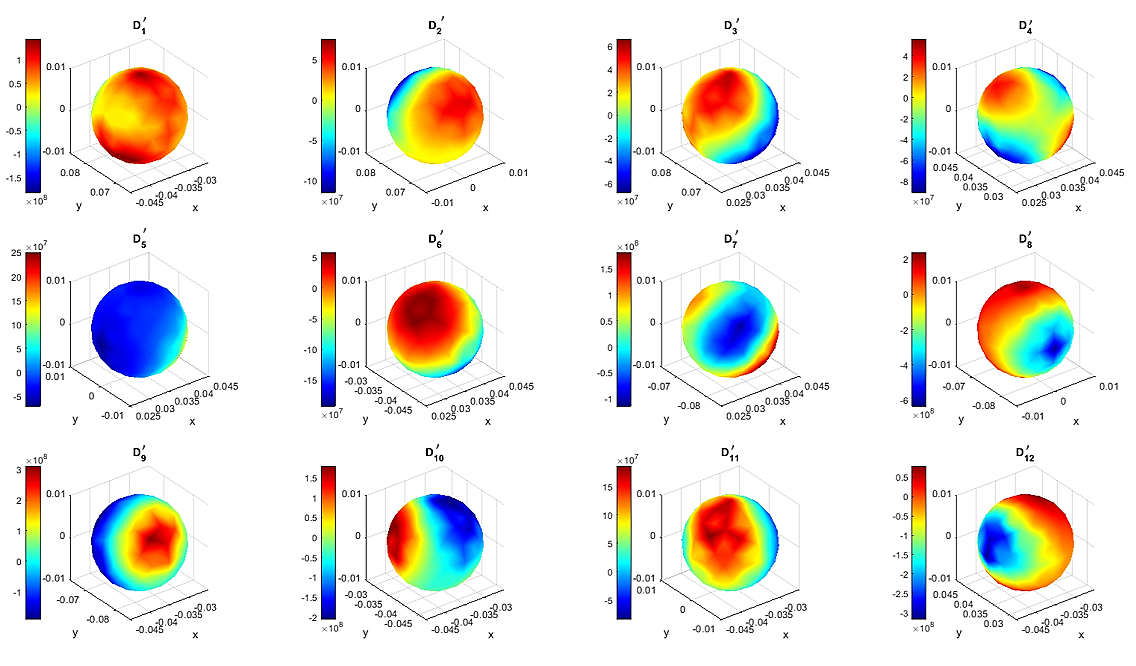}
	\caption{Real part of the computed density on the fictitious sources}
	\label{PhoneSimDensity}
\end{figure}

\begin{figure}[!h]
	\centering
	\includegraphics[width= \textwidth]{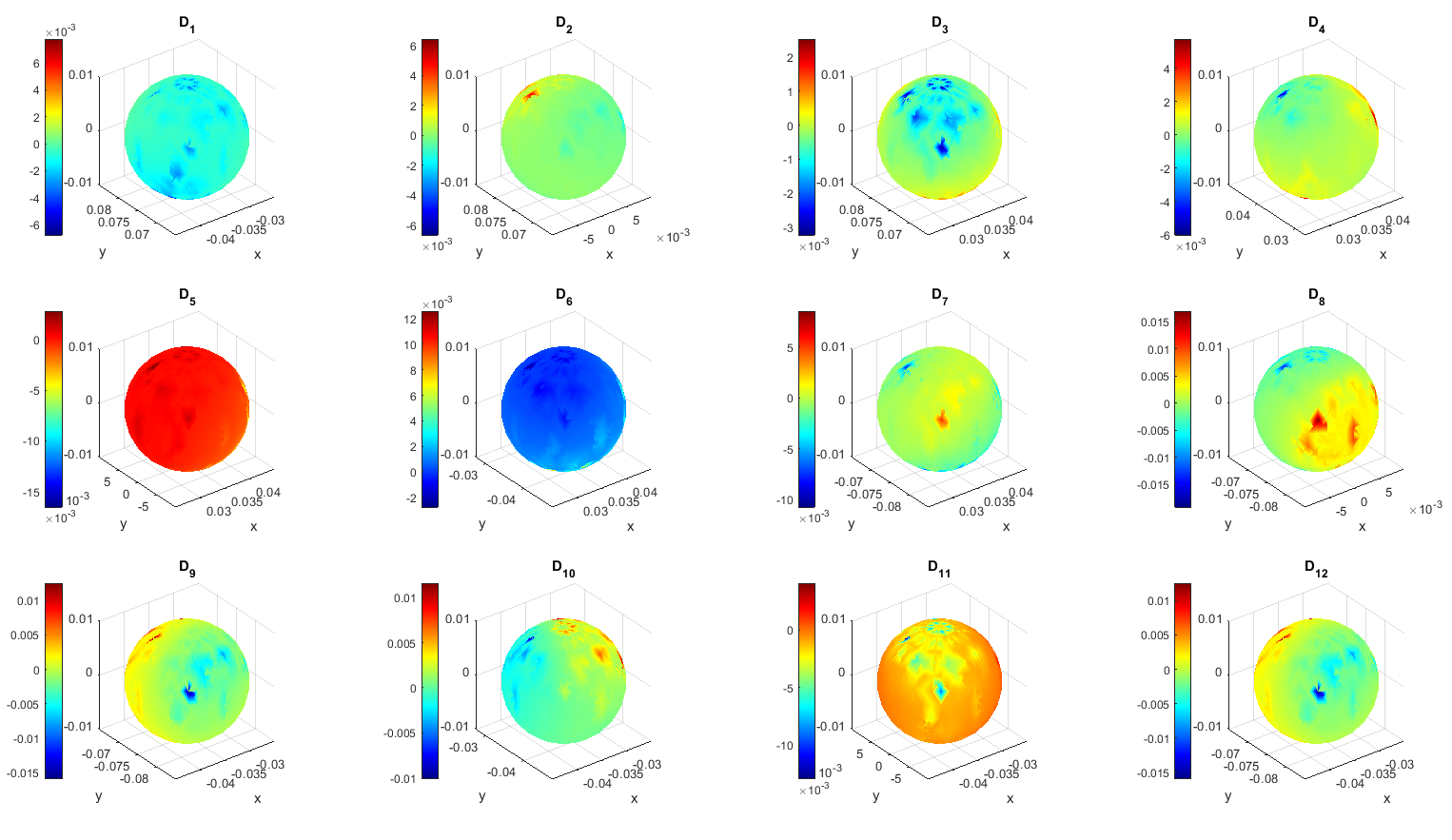}
	\caption{Real part of the corresponding normal velocity on the physical sources}
	\label{PhoneSimNV}
\end{figure}

The computed density from the numerical scheme and the resulting normal velocity on each source are shown in Figure \ref{PhoneSimDensity} and Figure \ref{PhoneSimNV}, respectively. Both show complex patterns with oscillations that may be difficult for physical instantiation.

\subsubsection{Time-domain Simulations: Shielded Localized Communications}

Next we employ again the Fourier synthesis developed in the beginning of Section \ref{TimeDomain} to extend the above discussion to the time-domain. Thus, we run a multitude of tests in each of which we impose the planewave $f_q(\Bx) = e^{i \Bx \cdot (k_q  \left < 0, 1, 0 \right >)}$ on the near control with $k_q = \overline {1, 50}$ (corresponding to frequencies from around 54.59 Hz to 2.73 KHz)  and a null field on the far control. The time-dynamic prescribed field $f$ and generated field $u$ are then obtained by superimposing the fields  from the different frequencies scaled by the time dependent factor $e^{-ik_qct}$, i.e., 

\[f(\Bx, t) = \sum_{q=1}^{50} f_q (\Bx) e^{-ik_qct} \text{ and } u(\Bx, t) = \sum_{q=1}^{50} u_q (\Bx) e^{-ik_qct} \] where $c = 343$ m/s is the assumed speed of sound in the medium. We let $t \in [0, 0.02]$ so that the time-domain animations will show around one period of the generated wave.

\begin{figure}[!h]
	\centering
	\includegraphics[width=0.8\textwidth]{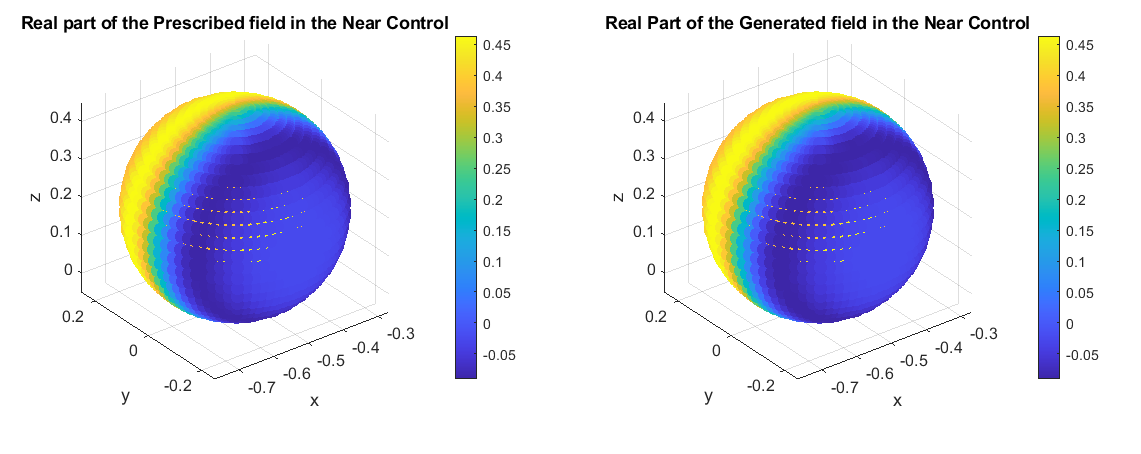}
	\caption{A visual comparison of the time-averaged real part of the prescribed and generated fields on the near control region}
	\label{PhoneSimTimeAveNear}
\end{figure}

\begin{figure}[!h]
	\centering
	\includegraphics[width=0.5 \textwidth]{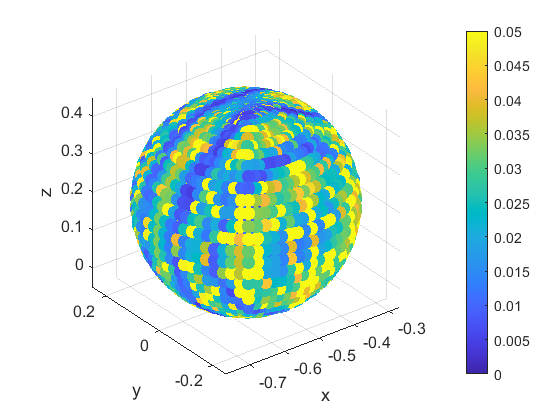}
	\caption{Plot of the time-averaged pointwise relative error}
	\label{PhoneSimTimeAveRelErr}
\end{figure}

The results of the time-domain synthesis are shown in an animation found in \cite{PhoneSimTimeNear}. The time-averaged plots of the prescribed and generated fields are shown in Figure \ref{PhoneSimTimeAveNear}. These show a good visual match between the fields. The time-averaged relative error is around around 3.87 \%. The plot of the time-averaged pointwise relative errors is shown in Figure \ref{PhoneSimTimeAveRelErr} while the time evolution of the pointwise relative error can be found in this animation \cite{PhoneSimTimeRelErr}. In these plots, the color map was truncated to show the variations in the relative error better.  

\begin{figure}[!h]
	\centering
	\includegraphics[width=0.5 \textwidth]{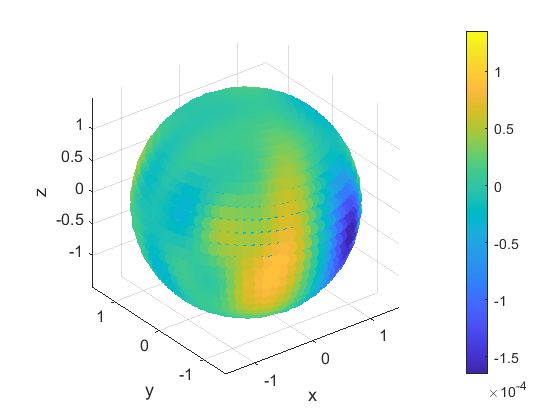}
	\caption{Plot of the time-averaged real part of the generated field in the surface of the null sphere}
	\label{PhoneSimTimeAveFar}
\end{figure}

The field synthesized on the surface of the  null sphere is also of small amplitude as seen in the time evolution animation found in \cite{PhoneSimTimeFar} and in the time-averaged plot in Figure \ref{PhoneSimTimeAveFar}. The supremum of the time-averaged field is around $1.74 \times 10^{-4}$. 

It is also of interest to observe the generated field on the cut plane $z = 0$ that cuts across the center of the sources. The complete time animation can be found in \cite{PhoneSimTimeCut0p0}. In this animation, the red circle shows the projection of the far control on $z = 0$, while the small black circles near the origin show the boundary of the sources. This show that the radiated energy is focused in the region near the sources and decays fast to achieve the near zero field on the far control.

\begin{figure}[!h]
	\centering
	\includegraphics[width= \textwidth]{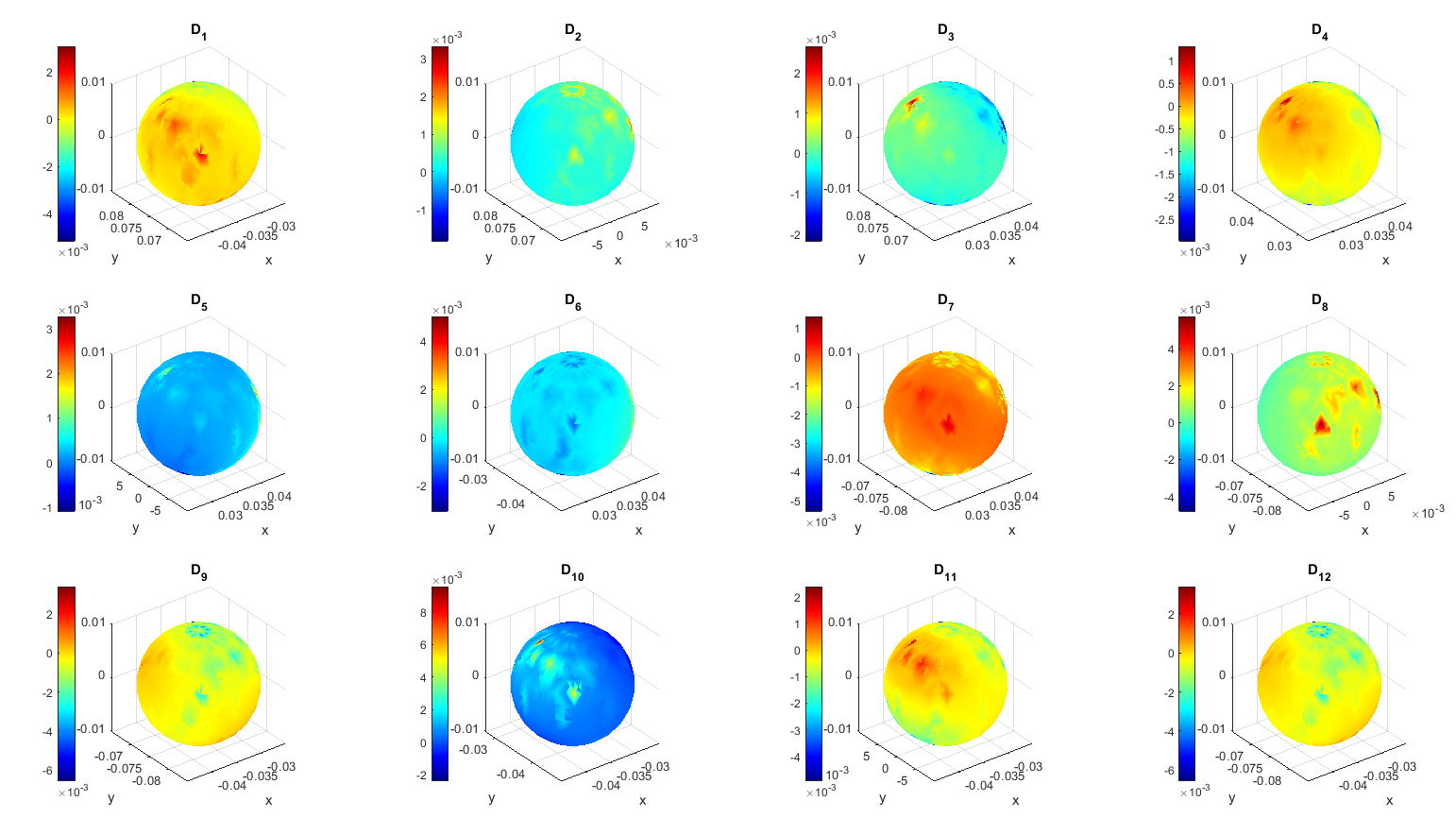}
	\caption{Plot of the time-averaged real part of the computed normal velocity}
	\label{PhoneSimTimeAveNV}
\end{figure}

Lastly, Figure \ref{PhoneSimTimeAveNV} shows time-averaged normal velocity, while its complete time evolution  can be found in the animation in \cite{PhoneSimTimeNV}. It can be noted that the complexity of the pattern and its oscillations in time may challenging for physical instantiation.

\section{Conclusion and future works}
\label{conclusion}
In this paper, we presented our strategy to achieve active control of acoustic fields in the context of two major applications: personal sound zones in bounded domains and shielded localized communications in free space. Our approach takes advantage of our previous works in the control of Helmholtz fields in free space using an almost non-radiating array of sources. We utilize the fact that we can design the desired source array by specifying a normal velocity (or surface pressure) on each source so that we can achieve desired fields in control regions near the sources and a zero field in the surface of a null sphere that contains the control regions and is well-separated from the region's boundary. The non-radiating character of the array ensures the fast decay of the generated field.  This is necessary so that the reverberations caused by the reflection of sound in the region's boundary will not significantly alter the primary field generated by the sources on the control region. This may overcome the need to model the region's boundary, which could possibly result to piecewise  mixed boundary conditions for the Helmholtz equation. 

The proposed approach was numerically implemented using a collocation method where the unknown density functions characterizing the normal velocities were represented using local basis functions. The resulting linear system was solved using Tikhonov regularization with the Morozov Discrepancy Principle to ensure the stability of the solution. Several numerical simulations were performed to illustrate the approach's utility in practical applications. The simulation results showed a successful synthesis of the desired fields on the control regions while maintaining a field with small amplitude in the null sphere. We also mention here that the theoretical and numerical results suggest that our time-domain active control approach will work well for any band-limited signal.

One pressing challenge posed by the obtained results is the complexity of the source feed inputs. Recent advances in the sound synthesis literature suggest the possibility of using heat nanotubes and very small transducers to instantiate the computed normal velocities (\cite{Hu2014, Lim2013, Brown2016, Asadzadeh2015, Qiao2020}). Other possible approaches to remedy this difficulty would be the use of different norms, such as the $L^1$ and TV norms in the regularization process. These alternative regularization methods may temper the oscillations produced in the current setup using the $L^2$ norm. 

Immediate extensions of this work include the performance of sensitivity tests with respect to variations in physically relevant problem parameters such as frequency, size of control regions, size and number of sources, distance between control regions and sources, relative position of control regions and sources, and the size of the null sphere and its distance from the region's boundary. One may also venture in the application of the proposed method in controlling acoustic waves in other environments such as in a homogeneous ocean and in an ocean of two homogeneous layers. In these cases, one may take advantage of the theoretical, functional, and numerical schemes developed in the works \cite{EgarguinIP2020, EgarguinWM2020, CQi2022, EgarguinIPSE2021}.

This work can also be extended to the control of electromagnetic waves in enclosed environments. Here, one can rely on the results established in \cite{Zeng2020, OEP2020, CQiEMLM, CQiEMSenst2022}. Specifically in \cite{OEP2020}, a scheme that uses the solution of the control problem of scalar Helmholtz fields to obatin smooth controls for the vectorial electromagnetic fields was developed utilizing Debeye potentials.

\section*{Acknowledgment}
NJA Egarguin's work towards this paper was funded by the University of the Philippines System Enhanced Creative Work and Research Grant (ECWRG-2022-1-8R).

\end {document}